\documentclass[doublecol,linenumbers]{epl2}

\usepackage{graphicx}
\usepackage{amsmath}

\usepackage{color}

\def\revision{}

\def\a{\alpha}

\def\ga{\gamma}

\def\phi{\varphi}
\def\la{\lambda}

\def\s{\sigma}

\def\om{\omega}

\def\R{{\bf R}}

\def\De{\Delta}

\def\pa{\partial}

\def\xb{{\bf x}}

\def\hb{{\bf h}}

\def\o+{\oplus}

\def\<{\langle}
\def\>{\rangle}

\def\Tr{\mathtt{Tr}}

\def\({\left(}
\def\){\right)}
\def\[{\left[}
\def\]{\right]}
\def\=#1{\bar #1}
\def\~#1{\widetilde #1}

\def\.#1{\dot #1}
\def\^#1{\widehat #1}
\def\"#1{\ddot #1}

\def\eeq{\end{equation}}
\def\beq{\begin{equation}}

\def\beql#1{\begin{equation} \label{#1}}

\def\eqref#1{(\ref{#1})}

\def\Fb{{\bf F}}

\def\salta#1{}

\title{On the isotropic-biaxial phase transition in nematic liquid
crystals}

\shorttitle{Isotropic-biaxial transition}

\author{G. Gaeta}

\institute{Dipartimento di Matematica, Universit\`a degli Studi di
Milano, via Saldini 50, I-20133 Milano (Italy)}


\abstract{ We apply a recently developed technique to determine
adapted coordinates for the sixth degree Landau-deGennes
potential, in which the potential is specially simple, to analyze
the possibility of a direct transition between the fully symmetric
state and a biaxial phase in nematic liquid crystals.
{Our results confirm, with simpler computations, results
by Allender and Longa.}
 }

\begin{document}

\maketitle

\section{Introduction}

Biaxial phases in nematic liquid crystals \cite{Fre,Fre2} receive
a continuing attention \cite{Bibook}. They have been approached
theoretically in many different ways (see e.g.
\cite{BZ0,SVD,BVGD,dMV,LGRV,Gry,KVP,GrL,DhS,To,Chi,Lon,BZ,Ste,ClT})
and sought for experimentally \cite{Bri,Luck}\footnote{{This
presented substantial difficulties: the first
experimental observation \cite{YuSa} came ten years after their
theoretical prediction \cite{Fre,Fre2}.}}, but many open questions
still exist \cite{Bibook}.\footnote{A large body of research has been devoted
to biaxial phases for liquid crystals made of molecules which are
biaxial themselves, or even have a more complex shape (e.g.
V-shaped, or tetrapodes \cite{Vij}). The biaxial phases considered
here could also be constituted by uniaxial molecules, as the order
tensor $Q$ (see below) represents the quadrupolar moment of the
molecular uniaxial distribution function.}

One of the widely debated questions concern the possibility of a
\emph{direct} transition from the fully symmetric state to a
stable biaxial phase, without an intermediate uniaxial phase
(generically, the biaxial phase appears in between the prolate and
the oblate uniaxial ones \cite{GLJ,AlL}). Even the simplest
theoretical analysis, based on the sixth degree Landau-deGennes
(LdG) potential \cite{deG,Vir} -- in the context of Landau theory
of phase transitions \cite{Lan1,Lan5} -- presents many
difficulties and intricacies \cite{GLJ,AlL}.


In this Letter, we want to apply a recently proposed method
\cite{AOP,Lan15a,Lan15b} for dealing with Landau theory along the
lines of (Poincar\'e-like) canonical perturbation theory
\cite{Arn,EMS1,Elp,CiW}, and in particular its modifications
(along the lines of ``further normalization'' \cite{CGs,RNF,Gae02}
in the dynamical systems parlance) to deal with problems involving
a phase transition \cite{Lan15b}.

The difficulty of the concrete problem at hand lies in that even
in this simple approach one has to analyze a sixth degree
potential depending on the leading parameter $\la$ and five
additional ones {(two of these could be eliminated by
suitably setting the scale for energy and $Q$; we won't do that)}.
That is, one is faced with a six-parameters family of sixth degree
Landau polynomials.

Our approach consists in separating the problem in two simpler
steps. The first and key step is the search for a ``normal form''
for this family: that is, a smaller family of (sixth degree)
Landau polynomials which, upon invertible maps, is in
correspondence with the full six-parameter family. Once this
smaller family -- concretely, a family depending on the leading
plus two additional parameters -- of simpler Landau polynomials,
referred to as the ``simplified LdG potential'', is obtained the
second step will consist in analyzing it and its critical points
with classical tools.

The main idea to implement the first and key step is borrowed from
Poincar\'e normal forms theory\footnote{Application of the
Poincar\'e theory also proved very effective in computing the
spectra of simple molecules \cite{Joy1,Joy2}.} for Dynamical
Systems \cite{Arn,EMS1,Elp}, and consists in using covariant
near-identity changes of coordinates \cite{AOP,Lan15a,Lan15b}.
This systematizes, using ideas by L. Michel \cite{Mic,MKZZ} and
other Authors \cite{AbS,CL,Cho}, an approach to simplification of
Landau potentials which appears to have been pioneered by Gufan
\cite{Guf}.

However, as we want to operate in a full neighborhood of the
transition point, we should pay attention these changes of
coordinates are well defined in all {of} such a region. An
abstract discussion of this point would require to introduce
``further normalization'' of Poincar\'e normal forms
\cite{CGs,RNF,Gae02}; however here we are concerned with a
specifical physical application, and all of our computations will
be completely explicit; it will thus suffice to check the
considered transformations are not singular in the {relevant}
region, and this will simply mean {that} no division by a
factor $\la$ should ever appear, where $\la$ is the parameter
{related to the quadratic part of the LdG potential,
hence controlling the local stability of the isotropic phase,
scaled so that} the critical value {for the local change
of stability is} $\la = 0$.

\section{Landau theory for nematic liquid crystals}

Nematic liquid crystals are described by a tensorial order
parameter $Q$ \cite{deG,Vir}; more precisely, this is a real
three-dimensional symmetric traceless matrix, hence parametrized
by five real numbers and which can be put in correspondence with a
five-dimensional vector $V = (z_1,...,z_5) \in \R^5$, \beql{eq:QV}
Q \ = \ \(
\begin{array}{ccc} z_1 & z_2 & z_3 \\ z_2 & z_4 & z_5 \\ z_3 & z_5
& - (z_1 + z_4) \end{array} \) \ . \eeq

The theory is covariant under the adjoint action of $SO(3)$ on
three-dimensional symmetric traceless matrices acting on $Q$; this
is also described in terms of the five-dimensional $SO(3)$
{representation} acting on $V$.

It is well known that this admits two algebraically independent
polynomial invariants, which hence also characterize orbits. A
convenient choice for these is just \beq T_2 \ = \
{(1/2)} \,  \Tr (Q^2) \ , \ \ T_3 \ = \ {(1/3)}
\, \Tr (Q^3 ) \ . \eeq In terms of the components $z_i$, these
read
 \begin{eqnarray}
T_2 &=& z_1^2  +  z_2^2  +  z_3^2  +  z_4^2  +  z_5^2  +  z_1  z_4 \ ; \label{eq:T23} \\
T_3 &=& z_1  (z_2^2 - z_4^2 - z_5^2)  -  z_4  (z_1^2 - z_2^2 +
z_3^2 )  +  2  z_2  z_3  z_5  \ . \nonumber
\end{eqnarray}
Here $T_2$ is related to $q = |Q| \revision{ = \sqrt{(1/2) \Tr (Q^T Q)}}$,
the amplitude of the order parameter $Q$, while $T_3$ is related to both $q$ and the measure of biaxiality $\om$; more precisely, we have \beql{eq:t2t3} T_2 \
= \ q^2 \ , \ \ T_3 \ = \ \frac{(1 - \om)}{\sqrt{6}} \, q^3 \ .
\eeq The inverse relations for \eqref{eq:t2t3} are of course
\beql{eq:qom} q \ = \ \sqrt{T_2} \ , \ \ \om \ = \ 1 \ - \sqrt{ 6
\, T_3^2 \, / \, T_2^3} ; \eeq {we take (the second of) these as
the definition of $\om$}.

We easily obtain {from \eqref{eq:T23} that} \beql{eq:Tomlimits}
{\Delta \ := \ } T_2^3 \ - \ 6 \, T_3^2 \ \ge \ 0 \ ,
\eeq which also entails \beql{eq:omlimits} 0 \ \le \ \om \ \le \ 1
\ . \eeq {Maximally biaxial states $\om = 1$ correspond
to $T_3 = 0$, while uniaxial states $\om = 0$ correspond to
$\Delta = 0$, i.e. to the boundary of the allowed region in the
$(T_2,T_3)$ plane.}

When describing nematic liquid crystals in terms of Landau theory
\cite{Lan1,Lan5} one uses the Landau-deGennes (LdG in the
following) potential \cite{deG}; on general grounds this is the
most general invariant (under the adjoint SO(3) action) sixth
order potential, and is therefore written as
\begin{eqnarray} \Phi &=& c_1 \,
T_2 \, + \, c_2 \, T_3 \, + \, c_3 \, T_2^2 \, + \, c_4 \, T_2 \,
T_3 \nonumber \\ & & \, + \, c_5 \, T_2^3 \, + \, c_6 \, T_3^2 \ .
\label{eq:Lan6}  \end{eqnarray} Here the $c_i$ are real
parameters, generally depending on the physical parameters
(temperature, pressure, etc); {we require (for stability)
$c_5
>0$, $c_5 + \revision{6} c_6 > 0$}. The state of the system is described by
minima\footnote{{One can check there are four nontrivial
branches -- i.e. $q \not= 0$ -- of critical points for $\Phi$;
some of these might be un-physical, i.e. with $q < 0$, depending
also on the parameter values.}} of $\Phi$. Obviously for $c_1
> 0$ the fully symmetric state $q=0$ is {locally} stable,
while this becomes {locally} unstable for $c_1 < 0$.

We will thus write $c_1 = - \la$, and consider $\la$ as the
leading parameter in the main transition, taking place for $\la =
0$ as $\la$ is varied. We will assume that the other parameters
$c_i$ are not varied; or at least that their variation is not
relevant in the considered region and can be disregarded. Note
this requires they are not zero at the transition, i.e. we are not
in a multi-critical case.

For $\la > 0$ {the isotropic state is locally unstable;}
the full rotational symmetry is spontaneously broken, and the
liquid crystals can in particular show uniaxial ($\om = 0$) or
biaxial ($\om \not= 0$) states. {These states can also be
present for $\la < 0$ and a locally stable isotropic state, due to
the appearance of lower energy configurations. Note that the
transition can also be (and necessarily is for $\la < 0$) first
order, i.e. the symmetry-breaking states can appear with non-zero
amplitude.}

A long standing question \cite{GLJ,AlL,Lon} is if there can be
stable branches of biaxial solutions branching off \emph{directly}
from the fully symmetric state $q=0$ at $\la = 0$. This
{matter} has been studied by several authors, but the
question is still open. An answer in the positive was provided by
Allender and Longa \cite{AlL} {in the frame of LdG
theory}, under rather restrictive conditions on the parameters.

\section{Change of variables; simplified LdG potential}

We want to simplify the LdG potential \eqref{eq:Lan6} by using a
\emph{near-identity covariant change of coordinates}. Here
near-identity means this will be of the form \beql{eq:CoV} z_i \ =
\ x_i \ + \ h_i (x) \ , \eeq with $h_i$ a nonlinear (polynomial)
function of the $x$; covariant means that we should preserve the
symmetry properties of the theory, and choose $\hb$ to transform
{in the same way} as $V$ under the $G=SO(3)$ action.
(Such a change of coordinates amounts to a nonlinear
reparametrization of the tensor $Q$.)

Covariant vector polynomials (for short, \emph{covariants}) under
this $G$ action are well known; they are generated by two basic
ones, i.e. a linear one, $$ \Fb_1 \ = \ \xb \ = \ ( x_1 , x_2 ,
x_3 , x_4 , x_5 )^T \ ; $$ and a quadratic one, \beql{eq:cov}
\Fb_2 \ = \ \(
\begin{array}{c}
(x_1^2 + x_2^2 + x_3^2 ) \, - \, 2 \, (x_1 x_4 + x_4^2 + x_5^2) \\
  3 (x_1 x_2 + x_2 x_4 + x_3 x_5) \\ 3 (x_2 x_5 - x_3 x_4 ) \\
    (x_2^2 + x_4^2 + x_5^2) \, - \, 2 \, (x_1^2 + x_3^2 + x_1 x_4)  \\
    3 (x_2 x_3 - x_1 x_5) \end{array} \) \ . \eeq
 The product of a (scalar) invariant and a (vector)
covariant is of course a covariant; thus the list of low order
covariant vectors is as follows:
$$ \Fb_1 = \xb, \ \Fb_2 , \ \Fb_3 = T_2 \xb ,\
 \Fb_4^{(1)} = T_3 \xb ,
 \ \Fb_4^{(2)} = T_2 \Fb_2
  \ . $$

We will {hence} consider a change of variables of the
form \eqref{eq:CoV}, with (it suffices to consider terms of degree
not higher than four) \beql{eq:covh} \hb \ = \ k_1 \, \Fb_2 \, +
\, k_2 \, \Fb_3 \, + \, k_3 \, \Fb_4^{(1)} \, + \, k_4 \,
\Fb_4^{(2)}
\ ; \eeq here the $k_i$ are arbitrary real constants, to be chosen
appropriately in a moment. It can be checked that with this, the
LdG potential $\Psi$ is changed into a potential of the same form,
i.e.
\begin{eqnarray} \Phi &=& c_1 \, T_2 \, + \, \ga_2 \, T_3 \, + \, \ga_3 \, T_2^2 \, + \,
\ga_4 \, T_2 \, T_3 \nonumber \\
& & \, + \, \ga_5 \, T_2^3 \, + \, \ga_6 \, T_3^2
\ + \ O(x^7) \ ; \label{eq:landauga}
 \end{eqnarray}
explicit expression for the coefficients $\ga_i$ can be easily
obtained with straightforward algebra. They become slightly
simpler by assuming, as we do in the following, $k_1 = 0$; in this
case they are given explicitly by
\begin{eqnarray*}
\ga_2 &=& c_2 \ , \\
\ga_3 &=& c_3 \ + \ 2 \, c_1 \, k_2 \ , \\
\ga_4 &=& c_4 \, + \, 3 \, c_2 \, k_2
  \, + \, 2 \, c_1 \, k_3 \, + \, 9 \, c_1 \, k_4 \ , \\
\ga_5 &=& {c_{5} \ + \ c_{1} \,
k_{2}^2 \ + \ 4 \, c_{3} \, k_{2} \ + \ 2 \, c_{2} \, k_{4} } \ , \\
\ga_6 &=& { c_{6} \ + \ 3 \, c_{2} \, k_{3} } \ .
\end{eqnarray*}

We would then like to choose the $k_i$ so to have as simple as
possible a potential $\Phi$. By this we mean one would like to
eliminate some of the terms (i.e. get some of the $\ga_i$ to
vanish). Note however that in order to guarantee thermodynamic
stability of the theory, $\Phi$ should be convex for large
$|\xb|$. A simple way to guarantee this is by having $\eta \
|\xb|^{2 k}$ (in this case with $k=3$), where $\eta$ is some
positive constant, as the highest order term.

We will work under the \emph{non-degeneracy assumption}
\beql{eq:c2not0} c_2 \ \not= \ 0 \ ; \eeq this means that at the
phase transition the next-to-leading order term is not vanishing.
In some of our considerations we will also assume, for the sake of
simplicity in the discussion, that $c_3 \not= 0$.

It is easily checked that requiring \beq \ga_4 \ = \ 0 \ , \ \
\ga_5 \ = \ 1 \ , \ \ \ga_6 \ = \ 1 \eeq admits a solution for
$k_2,k_3,k_4$ which is easily computed by an algebraic
manipulation program; full formulas are rather bulky and thus
omitted, but disregarding contributions of order $\la$ we get
\begin{eqnarray*}
k_1 &=& 0 \ , \\
k_2 &=& -\frac{c_4}{3 c_2} \ + \ O (\la ) \ , \\
k_3 &=& \frac{1 - c_6}{3 c_2} \ , \\
k_4 &=& \frac{3 c_2 + 4 c_3 c_4 - 3 c_2 c_5}{6 c_2^2} \ + \ O(\la
) \ . \end{eqnarray*}

With this choice we obtain
\begin{eqnarray*}
\ga_2 &=& c_2 \ , \\
\ga_3 &=& c_3 \ + \ [(2 c_4)/(3 c_2)] \, \la \ + \ O (\la^2 ) \ ,
\\
\ga_4 &=& 0 \ , \\
\ga_5 &=& 1 \ , \\
\ga_6 &=& 1 \ . \end{eqnarray*} The full expression for $\ga_3$ is
rather involved and not reported here; note that for $c_3 \not=0$
the sign of $\ga_3$ at small $\la$ is just that of $c_3$,
{while for $c_3=0$ it depends on the signs of $c_2$ and
$c_4$}.

Hence the LdG potential is reduced to \beql{eq:phiTP} \^\Phi \ = \
- \, \la \, T_2 \ + \ \ga_2 \, T_3 \ + \ \ga_3 \, T_2^2 \ + \
T_2^3 \ + \ T_3^2  \ . \eeq
 {In terms of the physical $(q,\om )$ variables this reads
$$ \^\Phi \ = \ - \la q^2 + \frac{\ga_2}{\sqrt{6}} (1 - \om) q^3 +
\ga_3 q^4 + \left[ 1 + \frac{1}{6} (1 - \om)^2 \right] q^6  \ . $$
}

Note that with this choice, and the notation introduced above, we
get $\eta = 1 + [(1 - \om)^2 /6] $; in view of \eqref{eq:omlimits}
this satisfies $ 1 \le \eta \le 7/6$.

It is maybe worth remarking that the reduced potential depends
essentially {on} $c_2$ and $c_3$, while dependence on
other parameters is rather weak and embodied in $\ga_3$. Recalling
that this simplified potential is valid in a neighborhood of the
transition point and for small $|q|$, the fact that only lower
order terns are relevant is certainly not a
surprise.\footnote{Degenerate (multi-critical) situations, with
\eqref{eq:c2not0} not holding or with some of the higher
coefficients vanishing at the transition point, can be analyzed
along the same lines.}

{It should be stressed that the nonlinear changes of
coordinates considered here produce higher order terms; in our
procedure we are only considering terms up to order six, i.e. we
are truncating the simplified LdG potential. This amounts to
considering a perturbation (small if working near zero) of the
original potential. It is in principles possible that the
perturbed (i.e. truncated) potential will display a qualitatively
different set of critical points than the original one; but if
this happens it means that the original LdG potential was
\emph{not structurally stable}. This appears not to be the case
here.}

\section{{Study of the simplified LdG potential \eqref{eq:phiTP}}}

We should now study the potential \eqref{eq:phiTP}, and in
particular its minima. It is convenient to perform this study in
orbit space, i.e. directly with the coordinates $T_2$, $T_3$ \cite{Mic,MKZZ}.

Critical points are identified as solutions to
\begin{eqnarray}
\frac{\pa \Phi}{\pa T_2} &=& - \, \la \ + \ 2 \, \ga_3 \, T_2 \ + \
3 \, T_2^2 \ = \ 0 \nonumber \\
\frac{\pa \Phi}{\pa T_3} &=& \ga_2 \ + \ 2 \, T_3 \ = \ 0 \ ;
\end{eqnarray}
these yield immediately\footnote{Note we are missing the trivial
solution $q=0$; this is due to the change of variable
\eqref{eq:t2t3} or \eqref{eq:qom} being singular at $q=0$.}
\beql{eq:T2T3simp} T_2 \ = \ \frac13 \, \( - \ga_3 \, + \, \a  \,
\sqrt{\ga_3^2 (1 + \mu)} \) \ , \ \ T_3 \ = \ - \ga_2/2 \ ; \eeq
here and in the following $\a = \pm 1$ and we write
\salta{\footnote{It should be noted that $\ga_3$ depends on $\la$,
so that \eqref{eq:mu} is a nonlinear change of parameter; more
precisely, $\mu (\la) = 9 c_2 \la^2 / (3 c_2 c_3 + 2 c_4 \la)^2$.
Note also that $\mu = 0$ if and only if $\la=0$, and near $\la =
0$ the function $\mu (\la)$ is regular provided $c_3 \not= 0$,
with $\mu' (0) = 3/c_3^2 > 0$. In the following we will consider
$\ga_3$ as a constant; this is not a problem,in that our analysis
will show that the relevant phenomena take place away from $\la =
0$; thus fully considering the $\la$-dependence of $\ga_3$ will
just produce a correction in the critical parameter values, not
any qualitative difference.}} \beql{eq:mu} \mu \ = \ 3 \, \la \ /
\ \ga_3^2 \ . \eeq

\begin{figure}
  \includegraphics[width=200pt]{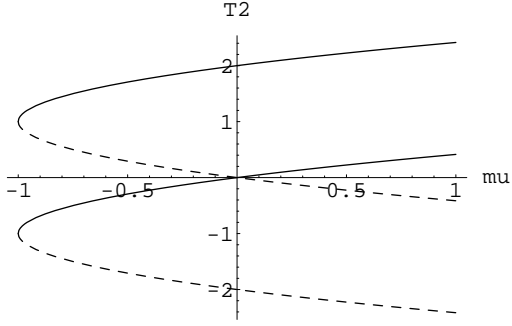}\\
  \caption{The value of $T_2$ as a function of $\mu$ along the solutions
  \eqref{eq:solsimp}, for $K=1$. Solid (dashed) curves refer to $\a=1$ ($\a=-1)$.
  The upper (lower) horizontal parabola is for $\s=-1$ ($\s=+1$).
  Only states with $T_2 \ge 0$ have physical meaning, due to $T_2 =q^2$.}\label{fig:0}
\end{figure}

The solutions \eqref{eq:T2T3simp} exist only for $\mu \ge -1$;
this will be understood without further notice from now on. It
should also be stressed that $T_2 = q^2$ requires $T_2 \ge 0$ for
the solutions to be physically relevant; see Fig.\ref{fig:0}.

Our discussion will depend on the signs of $\a$ and of $\ga_3$; it
is thus convenient to write $ \ga_3 =\s g$ with $g = |\ga_3 | \ge
0$ and $\s = \pm 1$. It will also be convenient to write $g = 3
K^2$, i.e. $ \ga_3 = 3 \s K^2 $ (say with $K >0$).

In this way the solutions \eqref{eq:T2T3simp} read simply
\beql{eq:solsimp} T_2 \ = \ K^2 \ \( - \s + \a \sqrt{1 + \mu} \) \
, \ \ T_3 \ = \ - \ga_2 / 2 \ . \eeq {We thus have again -- as for
the original LdG potential -- four nontrivial branches, indexed by
the signs of $\a$ and $\s$; some of these might be un-physical ($q
<0$), also depending on the parameters values; see Fig.\ref{fig:0}
and the discussion below.}

Note that $d T_2 / d \mu = \a [K^2/(2 \sqrt{1 + \mu})]$, i.e.
$T_2$ is strictly increasing (decreasing) with $\mu$ for $\a =1$
(for $\a = -1$).

The local stability of the solutions \eqref{eq:T2T3simp},
\eqref{eq:solsimp} can be simply analyzed by considering the
Hessian $H$ for the potential \eqref{eq:phiTP} at these solutions.
With trivial computations, $H$ and its eigenvalues $\zeta_i$ are
given by
\beq H \ = \ \begin{pmatrix} 2 \ga_3 + 6 T_2 &  0 \\
0 &  2 \end{pmatrix} \ ; \ \ \zeta_1 = 2 , \ \zeta_2 = 2 (\ga_3 +
3 T_2) \ ; \eeq thus along the solutions \eqref{eq:T2T3simp} we
have \beql{eq:EVH} \zeta_2 \ = \ \a \, \sqrt{ \ga_3^2 \, (1 +
\mu)} \ = \ 3 \, \a \, K^2 \, \sqrt{1 + \mu} \ . \eeq In
conclusion, the solutions with $\a = 1$ ($\a = - 1$) are always
locally stable (unstable).

We thus have four different branches of nontrivial solutions,
indexed by the signs of $\a$ and $\s$; these will be denoted as
$S_{(\pm \pm)}$. As mentioned above, $T_2 = q^2$ should be
positive to have physical meaning; it is immediately seen that for
$\mu >0$ only $\a = +1$ is allowed. Similarly, for $\mu < 0$, only
$\s = - 1$ is allowed. Thus $S_{(+ -)}$ is allowed for all $\mu >
-1$, while $S_{(+ +)}$ only for $\mu \ge 0$ and $S_{(- -)}$ only
for $\mu \le 0$; the latter one is locally unstable and thus not
of physical interest.

Let us now consider the $(q,\om)$ variables. It follows
immediately from $q = \sqrt{T_2}$ that \beql{eq:qsimp} q \ = \ K \
\sqrt{- \s \ + \ \a \sqrt{1 + \mu}} \ ; \eeq this produces, once
the signs of $\a$ and $\s$ are chosen, a ``universal'' behavior
for \beql{eq:chi} \chi \ := \ q/K \ \ = \ \sqrt{- \s \ + \ \a
\sqrt{1 + \mu}} \ . \eeq In the limit $\mu \to 0$ we get $q \to K
\sqrt{\a - \s }$; thus we have solutions branching off the
fully isotropic one for $\a = \s$. We will also have
solutions which are \emph{not} branching off the fully isotropic
one, corresponding to $\s = - \a$ as in $S_{(+-)}$.

As for $\om$, it follows from \eqref{eq:qom} that \beql{eq:omsimp}
\om \ = \ 1 \ - \ \frac{|\ga_2|}{\a \ K^3} \ \sqrt{\frac32 } \ \(
\frac{1}{\sqrt{1 + \mu} \, - \, \s  } \)^{3/2} \ . \eeq For given
$\a$ and $\s$ we have a ``universal'' behavior for \beql{eq:theta}
\theta \ := \ \sqrt{\frac23} \, \frac{K^3}{|\ga_2|} \ (1 - \om ) \
= \ \frac{1}{\a}  \ \( \frac{1}{\sqrt{1 + \mu} \, - \, \s  }
\)^{3/2} \ . \eeq It is immediately seen that if $\s=1$ then $\om$
is singular for $\mu \to 0$, so that \eqref{eq:omlimits} is surely
not satisfied at the branching point $\mu = 0$ (it might be
satisfied for larger values of $|\mu|$). For $\s = -1$ the
requirement \eqref{eq:omlimits} can be satisfied for $\mu \to 0$
provided $\a=1$ and depending on the values of $\ga_2$ and of $K$,
i.e. for $4  K^3 \ge |\ga_2| \sqrt{3}$. Recalling $K =
\sqrt{|\ga_3|/3}$ this reads \beql{eq:SMcond} 4 \ |\ga_3|^{3/2} \
\ge \ 9 \ |\ga_2| \ . \eeq

The situation is summarized in the following table; here ``loc
stab'' stands for locally stable, and $(q_0,\om_0)$ is the limit
of $(q,\om)$ for $\mu \to 0$.

\bigskip
\begin{tabular}{||c||c|c|c||}
\hline
   &  loc stab & $q_0=0$ & $0 \le \om_0 \le 1$ \\
  \hline
  $S_{(++)}$ &  yes & yes & no \\
  $S_{(+-)}$ &  yes & no  & see \eqref{eq:SMcond} \\
  $S_{(--)}$ &  no  & yes & no  \\
  \hline
\end{tabular}
\bigskip

It is clear from this table that no continuous (second order)
transition from the isotropic to a new stable state satisfying the
$\om$-condition \revision{\eqref{eq:omlimits}, hence \eqref{eq:SMcond},} is allowed at $\la=0$. On the
other hand, other types of transitions could and will be possible.

We stress that the above table considers the $\om$-condition
\eqref{eq:omlimits} only at the branching point $\mu = 0$. It is
well possible that a solution does not fulfill it at $\mu = 0$ but
complies with it at larger values of $|\mu|$.

For general values of $\mu$, the condition \eqref{eq:omlimits} is
better studied via the equivalent formulation
\eqref{eq:Tomlimits}.
On solutions this reads \beq \De \ = \ K^6 \ \( \a \,
\sqrt{1 + \mu} \, - \, \s \)^3 \ - \ (3/2) \, \ga_2^2 \ ; \eeq
thus \eqref{eq:Tomlimits} requires \beq \( \a \, \sqrt{1 + \mu} \
- \ \s \)^{3} \ \ge \ \frac32 \ \frac{\ga_2^2}{K^6} \ = \
\frac{81}{2} \ \frac{\ga_2^2}{|\ga_3 |^3} \ . \eeq

\begin{figure}
  \includegraphics[width=200pt]{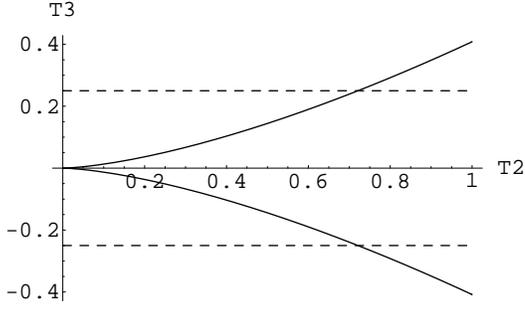}\\
  \caption{Solid lines represent the boundary of the region satisfying \eqref{eq:Tomlimits},
  i.e. \eqref{eq:omlimits}; on the boundary of this region,
  $\om=0$ (on the line $T_3=0$, $\om = 1$).
  The lines $T_3 = \pm \ga_2/2$ are in the region for $T_2 \ge L$.
  Here $\ga_2 = 1/2$, which by \eqref{eq:L} yields $L \simeq 0.72$.}\label{fig:1}
\end{figure}

Looking back at \eqref{eq:Tomlimits}, it is clear that the
$\om$-condition is satisfied for $|T_3| \le \sqrt{T_2^3/6}$ (with
$\om = 0$ on the boundary of this region); on the other hand, we
know that $T_2$ is monotone with $\mu$, and that on solutions $T_3
= - \ga_2/2$. Thus (see fig.\ref{fig:1}) 
the $\om$-condition is satisfied along the
solution for \beql{eq:L} | T_2 | \ \ge \ L \ := \ \left[ (3/2) \
\ga_2^2 \right]^{1/3} \ . \eeq The condition \eqref{eq:L} will
provide different conditions on the values of $\mu$ for different
solutions, i.e. for different signs of $\a$ and $\s$. At a given
$\mu$, the stable physical state will be the one with the lowest
energy among those satisfying the $\om$-condition.

In view of \eqref{eq:L} the limit value $\mu^*_\pm$ (which will be
a lower limit for $\a=1$) of $\mu$ for the $\om$-condition to be
satisfied on the solution $S_{(+ \pm)}$ is\footnote{In case it
results $\mu^*_+ < 0$, this should be meant as $\mu^*_+ = 0$, as
$S_{(++)}$ is only meaningful for $\mu \ge 0$; similarly, if
$\mu^*_- < -1$, this should be meant as $\mu^*_- = -1$.}
\beql{eq:mu0pm} \mu^*_\s \ = \ \frac{L \, (L + 2 \s K^2)}{K^4} \ .
\eeq Note that $\mu^*_+ > 0$, and it is always $\mu^*_- <
\mu^*_+$.

We should now evaluate the energy of states corresponding to the
solutions $S_{(+ \pm)}$. On these the simplified potential
\eqref{eq:phiTP} is trivial for $K=0$, and for $K \not= 0$ it
reads \beql{eq:phisol} \Phi_\s \ = \ K^6 \ \left[ (2 + 3 \mu) \,
\s \ - \ 2 \, \sqrt{(1 + \mu)^3} \ - \ \frac{\ga_2^2}{4 K^6}
\right] \ . \eeq Note that $\Phi_+$ is defined for $\mu \ge 0$,
while $\Phi_-$ for $\mu \ge -1$; when both of them are defined
($\mu \ge 0$), then $\Phi_+ - \Phi_- = 2 K^6 (2 + 3 \mu) > 0$.
Thus (recalling also $\mu^*_- < \mu^*_+$) the competition for the
lower energy is always between the isotropic state $S_0$, with
energy $\Phi_0 = 0$, and the state identified by $S_-$ with energy
\beq \Phi_- \ = \ - \, K^6 \ \left[ \ga_2^2/(4 K^6) \ + \  \left(
2 + 3 \mu + 2 (1+\mu)^{3/2} \right) \right] \ ; \eeq for $\mu >0$
it is always $\Phi_- < \Phi_0 = 0$, but it is clear this will also
hold for some range of negative $\mu$, i.e. for $\mu > \mu_0$ with
$\mu_0 < 0$.

We conclude that the state described by the solution $S_{(+-)}$
exists for $\mu \ge \mu^*_-$, and is stable for $\mu > \max
(\mu^*_- , \mu_0 ) $. As seen above, both $\mu^*_-$ and $\mu_0$
are strictly negative; thus there is \emph{always} a range of
negative $\mu$ (hence negative $\la$) for which the symmetry
breaking solution is stable.

If $\mu^*_- > \mu_0$, the symmetry breaking solution will appear
with $\om = 0$ and will then grow into a solution with $\om \not=
0$; but if instead $\mu^*_- < \mu_0$, the symmetry breaking
solution will appear with $\om \not= 0$. That is, in this case we
have a direct transition to a biaxial phase.

We would of course like to have some information about the range
of parameters allowing for such a situation. Noting that \beq
\frac{d \Phi_-}{d \mu} \ = \ - \ K^6 \ \( 1 \ + \ \sqrt{1 + \mu}
\) \ < \ 0 \ , \eeq it suffices to investigate if $\Phi_* :=
\Phi_- ( \mu^*_- )$ is positive (in which case $\mu_0 > \mu^*_-$)
or negative (in which case $\mu_0 < \mu^*_-$).

\begin{figure}
  \includegraphics[width=200pt]{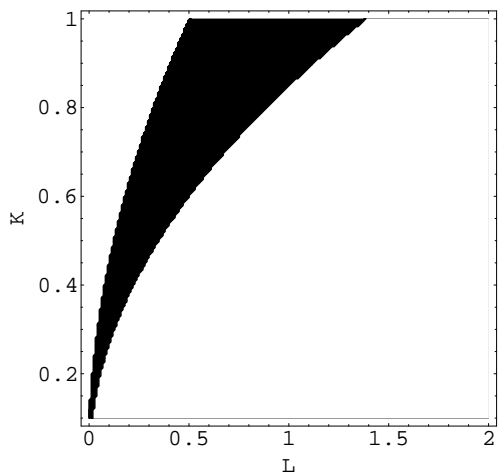}\\
  \caption{The black region in the parameter space is the one where $\Phi_*$, defined in \eqref{eq:Phi*},
  is positive.}\label{fig:2}
\end{figure}

Writing $\eta := |(L-K^2)/K^2|$, it results \beql{eq:Phi*} \Phi_*
\ = \ - \frac{L^3}{6} - 2  (1 + \eta)  K^6  -  L^2  (3 + 2 \eta)
K^2  +  2  L  (3 + 2 \eta)  K^4 \ . \eeq A numerical investigation
(see Fig.\ref{fig:2}) shows that there is a range of parameters
for which this is positive.

\section{Conclusions}

Considering Poincar\'e-like changes of coordinates holding in a
full neighborhood of $\la = 0$, the LdG potential of degree six,
{depending on six parameters}, can be led to a
{form involving} only three parameters.

The dependence on a smaller number of parameters allows to perform
easily a perturbation analysis near $\la = 0$, at least under the
non-degeneracy assumption \eqref{eq:c2not0}; this amounts to
requiring that the next-to-leading order term does not vanish
together with the leading order term and is thus a natural -- and
generically satisfied -- condition.

We have showed that under this hypothesis -- and within the sixth
degree Landau-deGennes theory -- {when passing to the simplified
LdG potential} there is no stable biaxial solution branching off
directly from the fully symmetric state via a second order
transition. {On the other hand, depending on the relation
between the two critical parameters $\mu^*_-$ and $\mu_0$ defined
above, there can be a direct first order transition from the fully
isotropic to a biaxial phase.}

This result should be compared with the one reported by Allender
and Longa \cite{AlL}; {they found that a stable biaxial
phase is present, and a direct transition from the fully isotropic
state to this phase is possible, for certain values of the
parameters. Thus our approach obtains the same qualitative results
as the standard one in this case.}

\begin{acknowledgements}
This work originated in the Program ``Mathematics of Liquid
Crystals'' at the Newton Institute. My research is partially
supported by MIUR-PRIN program under project 2010-JJ4KPA. {I would
like to thank the referees for interesting and very useful
constructive remarks.}
\end{acknowledgements}

\end{document}